\documentclass[prc,twocolumn,showpacs,preprintnumbers,amsmath,amssymb]{revtex4-1}
\usepackage[dvips]{graphicx}
\usepackage{dcolumn}
\usepackage{bm}

\begin{document}
\title{Is perturbative study of ground-state correlations valid?}
\author{Mitsuru Tohyama}
\affiliation{Faculty of Medicine, Kyorin University, Mitaka, Tokyo
  181-8611, Japan }
\begin{abstract}
Perturbative approaches have often been used to include the effects of ground-state correlations in extended theories of the random-phase approximation. 
Validity of such approaches is investigated for a solvable model where comparison with exact solutions can be made.
It is pointed out that there is a case where perturbative approaches give good results in spite of the fact that interaction strength is far beyond a perturbative region.
As a realistic case, the occupation probabilities in $^{16}$O are also calculated in the perturbative approach and compared with the results in exact diagonalization method.
\end{abstract}

\maketitle
\section{Introduction}
Magnetic dipole transitions in doubly $LS$ closed-shell nuclei such as $^{16}$O and $^{40}$Ca would be forbidden 
if the Hartree-Fock (HF) theory gave good description of the ground states of these nuclei.
Contrary to the HF assumption, strong magnetic dipole transitions have been observed in $^{16}$O and $^{40}$Ca \cite{gross} and the importance of 
ground-state correlations (core excitations) has been pointed out \cite{gross,alex}. Obviously
the random phase approximation (RPA) based on the HF ground state cannot be applied to the magnetic dipole transitions in these nuclei.
Various attempts have been made for the extension of RPA to include the effects of ground-state correlations.
The renormalized RPA (rRPA) \cite{rowe1,rowe2} 
includes ground-state correlation effects via the fractional occupation probability $n_\alpha$ of a single-particle state $\alpha$, which plays a role in opening new particle (p)--p and hole (h)--h transitions.
Ground-state correlations not only bring the fractional occupation of the single-particle states
but also give a correlated two-body density matrix $C_2$. The self-consistent RPA (SCRPA) \cite{scrpa1,scrpa2} includes both $n_\alpha$ and $C_2$: $C_2$ modifies
the energies of p--h, h--h and p--p pairs and also p--h, p--p and h--h correlations.
In rRPA and SCRPA $n_\alpha$ is self-consistently determined by the one-body amplitudes which are given as the solutions of extended RPA equations: 
In SCRPA $C_2$ is also given by the one-body amplitudes. The response function formalism of refs. \cite{taka1,taka2} uses 
a perturbative approach (PA) to calculate $n_\alpha$ and $C_2$. 
The extended second RPA (ESRPA) \cite{nishi,srpa} implements $n_\alpha$ and $C_2$ in the second RPA equation in a way similar to the response function approach.
PA has also been used to calculate $n_\alpha$'s in $^{16}$O \cite{taka3} and the sum rule values for spin-isospin modes in $^{16}$O and $^{40}$Ca \cite{adachi}.
The extended RPA of refs.\cite{toh07,toh17} uses $n_\alpha$ and $C_2$ obtained from ground-state calculations based on the time-dependent density-matrix theory (TDDM) \cite{toh20}.

Here we focus on a controversy over the application of PA, which still seems unsolved to our knowledge.
The PA calculations \cite{taka2,nishi,srpa,taka3,adachi} give large deviation of $n_\alpha$'s from the HF values ($n_\alpha=1$ or 0) in closed-shell nuclei. 
TDDM calculations also show that $n_\alpha$'s in $^{16}$O \cite{toh07} and $^{40}$Ca \cite{toh17} deviate more than 10 \% from the HF values. 
These PA and TDDM calculations indicate that the ground states of doubly-closed shell nuclei are highly correlated. However,
the authors of ref.\cite{neck} have claimed that the PA calculations \cite{taka2,nishi,srpa,taka3,adachi} use the perturbative expression for $n_\alpha$ far
beyond the region where the assumption of PA is justified 
and that the effects of 
ground-state correlations are largely overestimated in PA. In this paper the validity of the PA expressions for $n_\alpha$ and $C_2$ is investigated in the case of the Lipkin model \cite{Lip}.
We use the TDDM equations to interpret the exact solutions because TDDM well reproduces the exact values.
The results for the ground state of $^{16}$O are also presented as a realistic case.
It is shown that there is a situation where the PA expressions give good results in spite of the fact that interaction strength is beyond a perturbative region.
The paper is organized as follows. The formulation of TDDM and its perturbative limit are presented in sect. 2,
the obtained results are given in sect. 3 and sect. 4 is devoted to summary.

\section{Formulation}
We use the TDDM equations to interpret the behavior of the exact solutions because TDDM well reproduces the exact values. The formulation of TDDM and its perturbative limit are given below.

\subsection{Time-dependent density-matrix theory}
The TDDM equations consist of
the coupled equations of motion for the one-body density matrix (the occupation matrix) $n_{\alpha\alpha'}$
and the correlated part of the two-body density matrix $C_{\alpha\beta\alpha'\beta'}$ ($C_2$) \cite{toh20}.
These matrices are defined as
\begin{eqnarray}
n_{\alpha\alpha'}(t)&=&\langle\Phi(t)|a^+_{\alpha'} a_\alpha|\Phi(t)\rangle,
\\
C_{\alpha\beta\alpha'\beta'}(t)&=&\rho_{\alpha\beta\alpha'\beta'}(t)
-(n_{\alpha\alpha'}(t)n_{\beta\beta'}(t)
\nonumber \\
&-&n_{\alpha\beta'}(t)n_{\beta\alpha'}(t)),
\label{rho2}
\end{eqnarray}
where $|\Phi(t)\rangle$ is the time-dependent total wavefunction
$|\Phi(t)\rangle=\exp[-iHt] |\Phi(t=0)\rangle$, $\rho_{\alpha\beta\alpha'\beta'}$ is the two-body density matrix (
$\rho_{\alpha\beta\alpha'\beta'}(t)=\langle\Phi(t)|a^+_{\alpha'}a^+_{\beta'}a_{\beta}a_{\alpha}|\Phi(t)\rangle$). Here, $H$ is the total Hamiltonian and
units $\hbar=1$ are used hereafter.
The equations of motion for $n_{\alpha\alpha'}$ and $C_{\alpha\beta\alpha'\beta'}$  are written as
\begin{eqnarray}
i \dot{n}_{\alpha\alpha'}&=&
\sum_{\lambda}(\epsilon_{\alpha}-\epsilon_{\alpha'}){n}_{\alpha\alpha'}
\nonumber \\
&+&\sum_{\lambda_1\lambda_2\lambda_3}
[\langle\alpha\lambda_1|v|\lambda_2\lambda_3\rangle C_{\lambda_2\lambda_3\alpha'\lambda_1}
\nonumber \\
&-&C_{\alpha\lambda_1\lambda_2\lambda_3}\langle\lambda_2\lambda_3|v|\alpha'\lambda_1\rangle],
\label{n}
\end{eqnarray}
\begin{eqnarray}
i\dot{C}_{\alpha\beta\alpha'\beta'}&=&
(\epsilon_{\alpha}+\epsilon_{\beta}
-\epsilon_{\alpha'}-\epsilon_{\beta'}){C}_{\alpha\beta\alpha'\beta'}
\nonumber \\
&+&B_{\alpha\beta\alpha'\beta'}+P_{\alpha\beta\alpha'\beta'}+H_{\alpha\beta\alpha'\beta'}+T_{\alpha\beta\alpha'\beta'},
\label{C2}
\end{eqnarray}
where $\epsilon_{\alpha}$ is the single-particle energy and $\langle\alpha\beta|v|\alpha'\beta'\rangle$ the matrix element of a two-body interaction $v$.
The term $B_{\alpha\beta\alpha'\beta'}$ in eq. (\ref{C2}) includes only the occupation matrices and describes  2 particle (p) -- 2 hole (h)
and 2h--2p
excitations, playing a role as a source term. The terms $P_{\alpha\beta\alpha'\beta'}$ and $H_{\alpha\beta\alpha'\beta'}$
contain $C_2$ and express
p--p (and h--h) and p--h
correlations to infinite order, respectively \cite{toh20}.
The $T_{\alpha\beta\alpha'\beta'}$ term gives the coupling to the three-body correlation matrix ($C_3$).
Approximations for $C_3$ are needed to close the equations of motion within $n_{\alpha\alpha'}$ and $C_2$.
In this study we use the following truncation scheme \cite{ts2017}
\begin{eqnarray}
C_{\rm p_1p_2h_1p_3p_4h_2}&=&\frac{1}{\mathcal N}\sum_{\rm h}C_{\rm hh_1p_3p_4}C_{\rm p_1p_2h_2h},
\label{purt11}\\
C_{\rm p_1h_1h_2p_2h_3h_4}&=&\frac{1}{\mathcal N}\sum_{\rm p}C_{\rm h_1h_2p_2p}C_{\rm p_1ph_3h_4},
\label{purt22}
\end{eqnarray}
where p and h refer to particle and hole states, respectively, and ${\mathcal N}$ is given by 
\begin{eqnarray}
{\mathcal N}&&=1+\frac{1}{4}\sum_{\rm pp'hh'}C_{\rm pp'hh'}C_{\rm hh'pp'}.
\label{norm}
\end{eqnarray}
These expressions were derived from perturbative consideration using
the Coupled-Cluster-Doubles (CCD)-like ground state wavefunction \cite{shavitt} $|Z\rangle$ 
\begin{eqnarray}
|Z\rangle=e^Z|{\rm HF}\rangle\approx (1+Z)|{\rm HF}\rangle
\end{eqnarray}
with 
\begin{eqnarray}
Z=\frac{1}{4}\sum_{\rm pp'hh'}z_{\rm pp'hh'}a^\dag_{\rm p}a^\dag_{\rm p'}a_{\rm h'}a_{\rm h},
\end{eqnarray}
where $|{\rm HF}\rangle$ is the HF ground state and in the lowest order of $v$  
$z_{\rm pp'hh'}$ is given by
\begin{eqnarray}
z_{\rm pp'hh'}&=&-\frac{\langle {\rm pp'}|v|{\rm hh'}\rangle_A}{\epsilon_{\rm p}+\epsilon_{\rm p'}
-\epsilon_{\rm h}-\epsilon_{\rm h'}}.
\label{pertz}
\end{eqnarray}
Here, the subscript $A$ means that the corresponding matrix is antisymmetrized. 
In the lowest order of $z_{\rm pp'hh'}$ 
the two-body correlation matrices are given by
\begin{eqnarray}
C_{\rm pp'hh'}&\approx& z_{\rm pp'hh'},
\label{cz}
\\
C_{\rm hh'pp'}&\approx& z^*_{\rm pp'hh'},
\end{eqnarray}
and the three-body correlation matrices by
\begin{eqnarray}
C_{\rm p_1p_2h_1p_3p_4h_2}&\approx&\sum_{\rm h}z^*_{\rm p_3p_4hh_1}z_{\rm p_1p_2h_2h},
\label{purt01}\\
C_{\rm p_1h_1h_2p_2h_3h_4}&\approx&\sum_{\rm p}z^*_{\rm p_2ph_1h_2}z_{\rm p_1ph_3h_4}.
\label{purt02}
\end{eqnarray}
These relations suggest the expressions for 
$C_{\alpha\beta\gamma\alpha'\beta'\gamma'}$
in terms of $C_{\rm pp'hh'}$ given by eqs. (\ref{purt11}) and (\ref{purt22}).
The factor ${\mathcal N}$ was introduced to simulate many-body effects which reduce $C_3$ in large $N$ systems and (or) strongly interacting regions of the Lipkin model.
In a perturbative region where the second term on the right-hand side of Equation (\ref{norm}) is smaller than unity, 
${\mathcal N}$ has the meaning of the normalization factor of the total wavefunction.

\subsection{Perturbative limit of TDDM}
In the lowest order of $v$ the TDDM equations give the perturbative expressions for $n_\alpha$ and $C_2$ \cite{st2016}.
For small $v$ the $B$ term is dominant in eq. (\ref{C2}) and the stationary condition of $C_2$ gives
\begin{eqnarray}
i\dot{C}_{\alpha\beta\alpha'\beta'}&=&
(\epsilon_{\alpha}
+\epsilon_{\beta}
-\epsilon_{\alpha'}
-\epsilon_{\beta'}){C}_{\alpha\beta\alpha'\beta'}
\nonumber \\
&+&B_{\alpha\beta\alpha'\beta'}=0,
\label{pert}
\end{eqnarray}
where $B_{\alpha\beta\alpha'\beta'}$ is given by 
\begin{eqnarray}
B_{\alpha\beta\alpha'\beta'}&=&\langle \alpha\beta|v|\alpha'\beta'\rangle_A ((1-n_\alpha)(1-n_\beta)n_{\alpha'}n_{\beta'}
\nonumber \\
&-&n_{\alpha}n_{\beta}(1-n_{\alpha'})(1-n_{\beta'})).
\label{B}
\end{eqnarray}
Here, the occupation matrix is assumed diagonal.
Using the HF assumption $n_\alpha=1$ or 0, we obtain the same result as eq. (\ref{pertz})
\begin{eqnarray}
C_{\rm pp'hh'}&=&-\frac{\langle {\rm pp'}|v|{\rm hh'}\rangle_A}{\epsilon_{\rm p}+\epsilon_{\rm p'}
-\epsilon_{\rm h}-\epsilon_{\rm h'}}.
\label{pertC}
\end{eqnarray}
Inserting this into eq. (\ref{n}) with $\dot{n}_{\alpha\alpha'}=0$,
we obtain
\begin{eqnarray}
n_{\rm p}&=&\lim_{{\rm p'}\rightarrow {\rm p}}n_{\rm{pp'}}=\lim_{{\rm p'}\rightarrow {\rm p}}\frac{1}{\epsilon_{\rm p'}-\epsilon_{\rm p}}
\nonumber \\
&\times&\sum_{\rm p_1h_1h_2}(\frac{\langle {\rm p p_1}|v|{\rm h_1h_2}\rangle \langle{\rm h_1h_2}|v|{\rm p'p_1}\rangle _A}{\epsilon_{\rm h_1}+\epsilon_{\rm h_2}-\epsilon_{\rm p'}-\epsilon_{\rm p_1}}
\nonumber \\
&+&\frac{\langle{\rm p p_1}|v|{\rm h_1h_2}\rangle _A}{\epsilon_{\rm p}+\epsilon_{\rm p_1}-\epsilon_{\rm h_1}-\epsilon_{\rm h_2}}\langle{\rm h_1h_2}|v|{\rm p'p_1}\rangle)
\nonumber \\
&=&\frac{1}{2}\sum_{\rm p_1h_1h_2}\left|\frac{\langle{\rm p p_1}|v|{\rm h_1h_2}\rangle _A}{\epsilon_{\rm p}+\epsilon_{\rm p_1}-\epsilon_{\rm h_1}-\epsilon_{\rm h_2}}\right|^2.
\label{pertp}
\end{eqnarray}
Similarly, $\Delta n_{\rm h}=1-n_{\rm h}$ is given as 
\begin{eqnarray}
\Delta n_{\rm h}=\frac{1}{2}\sum_{\rm h_1p_1p_2}\left|\frac{\langle{\rm h h_1}|v|{\rm p_1p_2}\rangle _A}{\epsilon_{\rm h}+\epsilon_{\rm h_1}-\epsilon_{\rm p_1}-\epsilon_{\rm p_2}}\right|^2.
\label{perth}
\end{eqnarray}
Equations (\ref{pertC})-(\ref{perth}) have been used in the PA calculations \cite{taka2,nishi,srpa,taka3,adachi}.
In the lowest order of $v$, ${\mathcal N}$ in eq. (\ref{norm}) is expressed as
\begin{eqnarray}
{\mathcal N}=1+\frac{1}{2}\left(\sum_{\rm h}\Delta n_{\rm h}+\sum_{\rm p} n_{\rm p}\right).
\end{eqnarray}
When $n_\alpha$'s deviate more than 10 \% 
from the HF values as the perturbative calculations \cite{taka2,nishi,srpa,taka3,adachi} have suggested, the perturbative condition $\mathcal N\approx 1$ is easily violated \cite{neck}.
In eq. (\ref{C2}) 
the $B$ term decreases with increasing $n_{\rm p}$ (and decreasing $n_{\rm h}$) due to the occupation factor $(1-n_{\rm p})(1-n_{\rm p'})n_{\rm h}n_{\rm h'}$
and the $P$, $H$ and $T$ terms as a whole play a role in modifying the energies of the 2p--2h configurations at least in an interaction region where $C_{\rm pp'hh'}$ still dominates $C_2$.
Therefore, the results in TDDM for $C_{\rm pp'hh'}$ in that interaction region
may be interpreted by an expression similar to eq. (\ref{pertC})
\begin{eqnarray}
C_{\rm pp'hh'}=-\frac{B_{\rm pp'hh'}}{E_{\rm pp'hh'}},
\label{BoE}
\end{eqnarray}
where $E_{\rm pp'hh'}=\epsilon_{\rm p}+\epsilon_{\rm p'} -\epsilon_{\rm h}-\epsilon_{\rm h'}+\Delta E$. Here, 
$\Delta E$ describes a decrease in the 2p--2h energy with increasing interaction strength. In the case of the Lipkin model
the value of $\Delta E$ can be estimated from the energy of the second excited state which has the same symmetry as the ground state.

\section{Results}
The Lipkin model \cite{Lip} has extensively been used to test theoretical models. It describes an $N$-fermions system with two
$N$-fold degenerate levels. The upper (lower) levels have energies $\epsilon/2$ ($-\epsilon/2$) and quantum number
$p$ ($-p$) with $p=1,2,...,N$.
The Hamiltonian is given by
\begin{equation}
{H}=\epsilon {J}_{z}+\frac{V}{2}({J}_+^2+{J}_-^2),
\label{elipkin}
\end{equation}
where the operators are the followings
\begin{eqnarray}
{J}_z&=&\frac{1}{2}\sum_{p=1}^N(c_p^{+}c_p-{c^+_{-p}}c_{-p}), \\
{J}_{+}&=&{J}_{-}^{+}=\sum_{p=1}^N c_p^{+}c_{-p}.
\end{eqnarray}
For $\chi=|V|(N-1)/\epsilon \le 1$
the HF ground state is given by 
$|{\rm HF}\rangle=\prod_{p=1}^Nc^+_{-p}|0\rangle$,
where $|0\rangle$ is the true vacuum.
For $\chi>1$ the lowest single-particle states are obtained by the transformation 
\begin{eqnarray}
\left(
\begin{array}{c}
a^+_{-p}\\
a^+_{p}
\end{array}
\right)=\left(
\begin{array}{cc}
\cos\alpha&\sin\alpha\\
-\sin\alpha&\cos\alpha
\end{array}
\right)
\left(
\begin{array}{c}
c^+_{-p}\\
c^+_{p}\\
\end{array}
\right),
\nonumber \\
\label{2l-trans}
\end{eqnarray}
where $\alpha$ satisfies $\cos2\alpha=1/\chi$. The HF ground state in this case is often called the 'deformed' HF (DHF) state and is given by 
$|{\rm DHF(\alpha)}\rangle=\prod_{p=1}^Na^+_{-p}|0\rangle$.

\begin{figure} 
\resizebox{0.5\textwidth}{!}{
\includegraphics{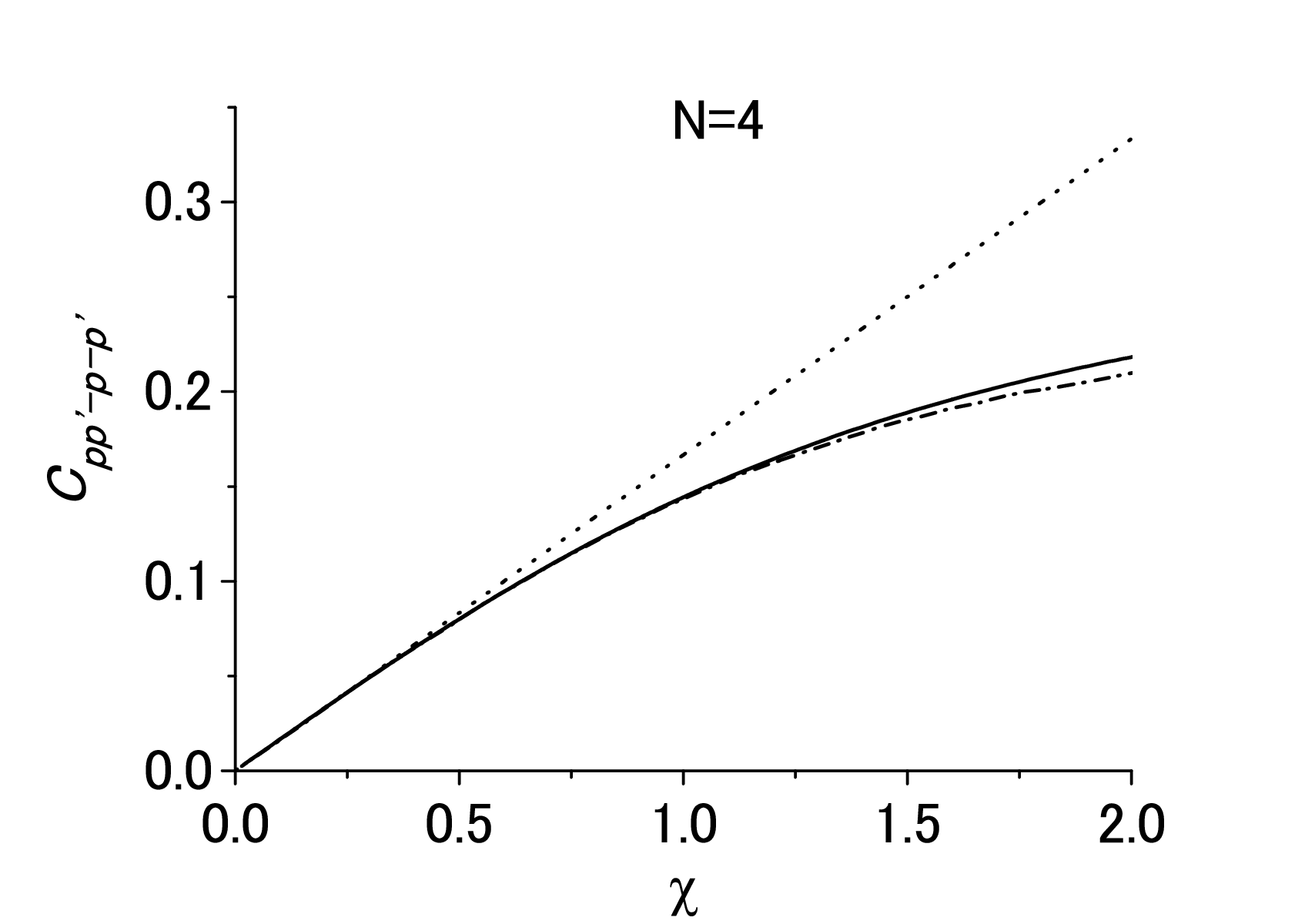}
}
\caption{$C_{pp'-p-p'}$ in PA (dotted line) as a function of $\chi=|V|(N-1)/\epsilon$ for $N=4$.
The results in TDDM and the exact values are shown with the dot-dashed and solid lines, respectively.} 
\label{pert4C} 
\end{figure} 
\begin{figure}
\resizebox{0.5\textwidth}{!}{ 
\includegraphics{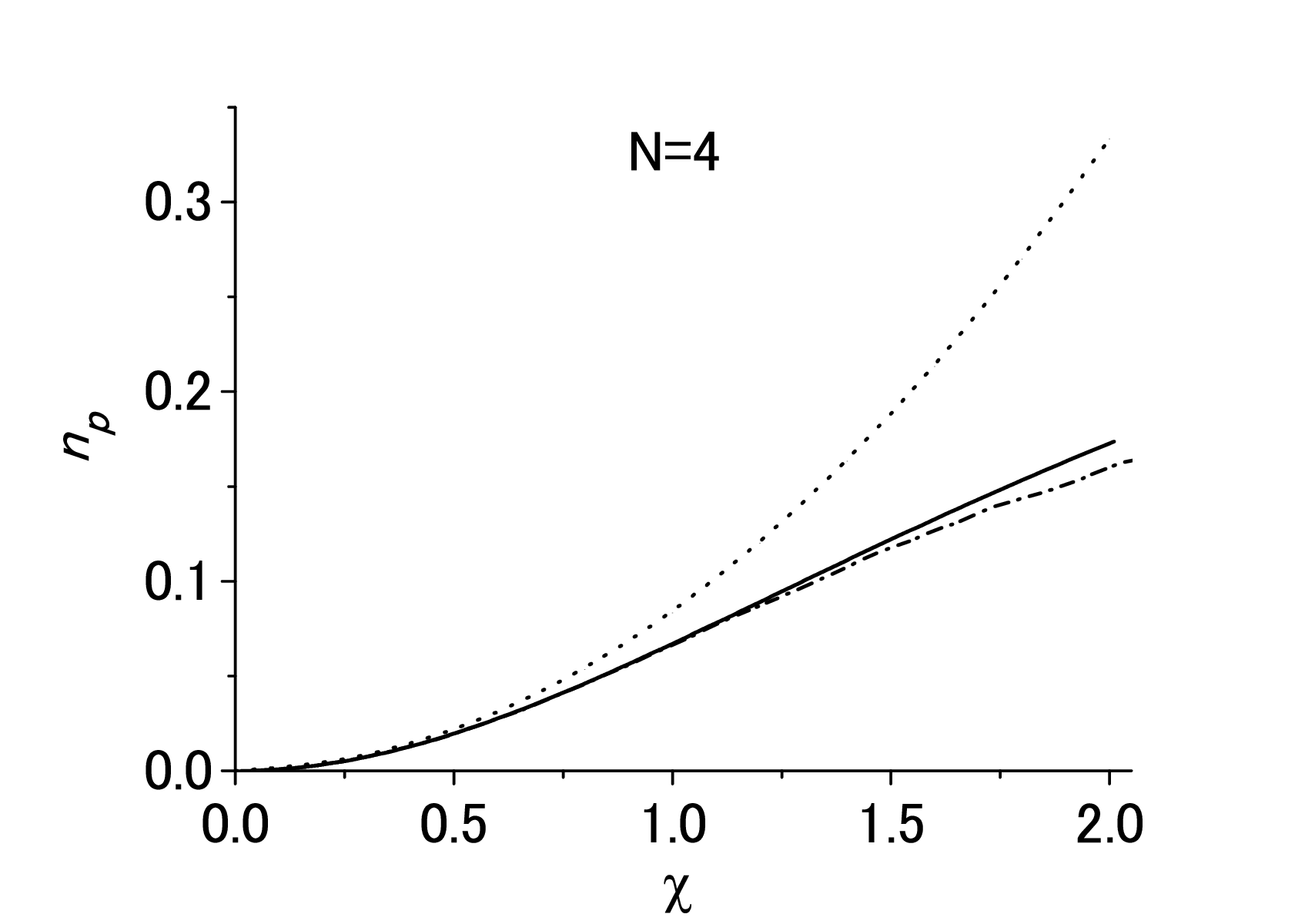}
}
\caption{$n_p$ of the upper state in PA (dotted line) as a function of $\chi$ for $N=4$.
The results in TDDM and the exact values are shown with the dot-dashed and solid lines, respectively.} 
\label{pert4n} 
\end{figure} 
\begin{figure} 
\resizebox{0.5\textwidth}{!}{
\includegraphics{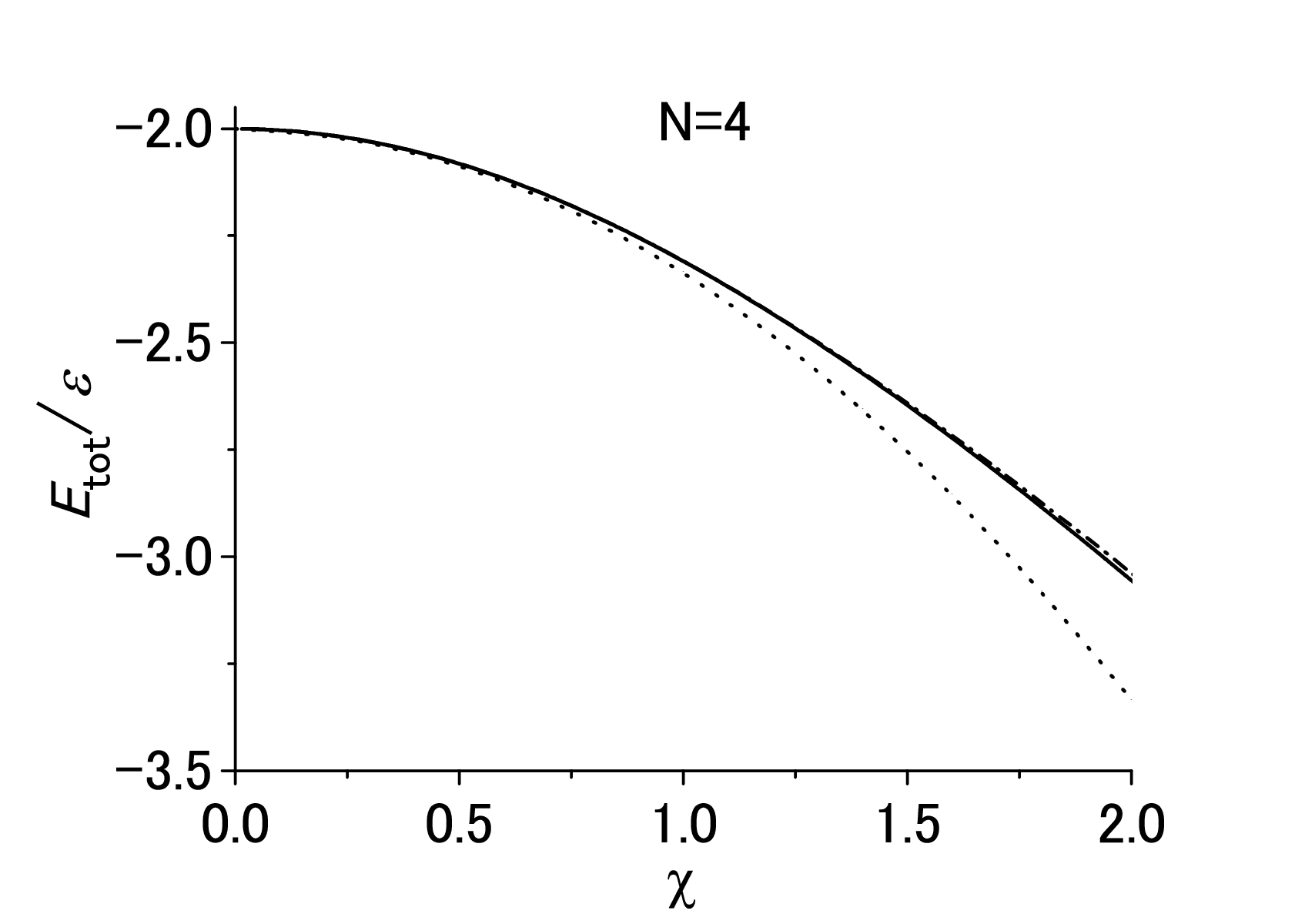}
}
\caption{$E_{\rm tot}$ in PA (dotted line) as a function of $\chi$ for $N=4$. 
The results in TDDM and the exact values are shown with the dot-dashed and solid lines, respectively.} 
\label{pert4E} 
\end{figure} 
The 2p--2h component $C_{pp'-p-p'}$ of $C_2$, the occupation probability $n_p$ of the upper state and the ground-state energy $E_{\rm tot}$ 
calculated in PA (dotted line) and TDDM (dot-dashed line) are shown, respectively, 
in figs. \ref{pert4C}--\ref{pert4E} as functions of $\chi$ for $N=4$. The TDDM equations eqs. (\ref{n}) and (\ref{C2}) are solved by using the adiabatic method \cite{ts2017}.
The solid lines depict the exact values. The TDDM results are in good agreement with the exact solutions.
The results in PA for $C_{pp'-p-p'}$ start to deviate from the exact values around $\chi=0.5$. Since the energy of the 2p--2h configuration estimated from the exact energy of the second excited state
is unchanged in the case of $N=4$,
the deviation of the PA results from the exact values is solely explained by the $B$ term in eq. (\ref{BoE}), that is, the decrease of the occupation factor $(1-n_p)(1-n_{p'})n_{-p}n_{-p'}$
with increasing $n_p$ and decreasing $n_{-p}$ . The ground-state energy $E_{\rm tot}$ in PA remains close to the exact values till $\chi=1$ where $n_p$ and $C_{pp'-p-p'}$ in PA deviate about 20 \% from the exact values.
This is due to the cancellation of errors in the single-particle part and the interaction part of $E_{\rm tot}$. As can be seen in figs. \ref{pert4C}-\ref{pert4E}, 
the perturbative expressions are valid for $\chi<0.5$
where PA gives ${\mathcal N}<1.077$, meaning the genuine perturbative region.
At $\chi=1$ the value of $n_p$ is $0.083$, the error of $n_p$ is 26 \% and ${\mathcal N}=1.333$. 
The error of PA to this extent might be tolerable.

\begin{figure} 
\resizebox{0.5\textwidth}{!}{ 
\includegraphics{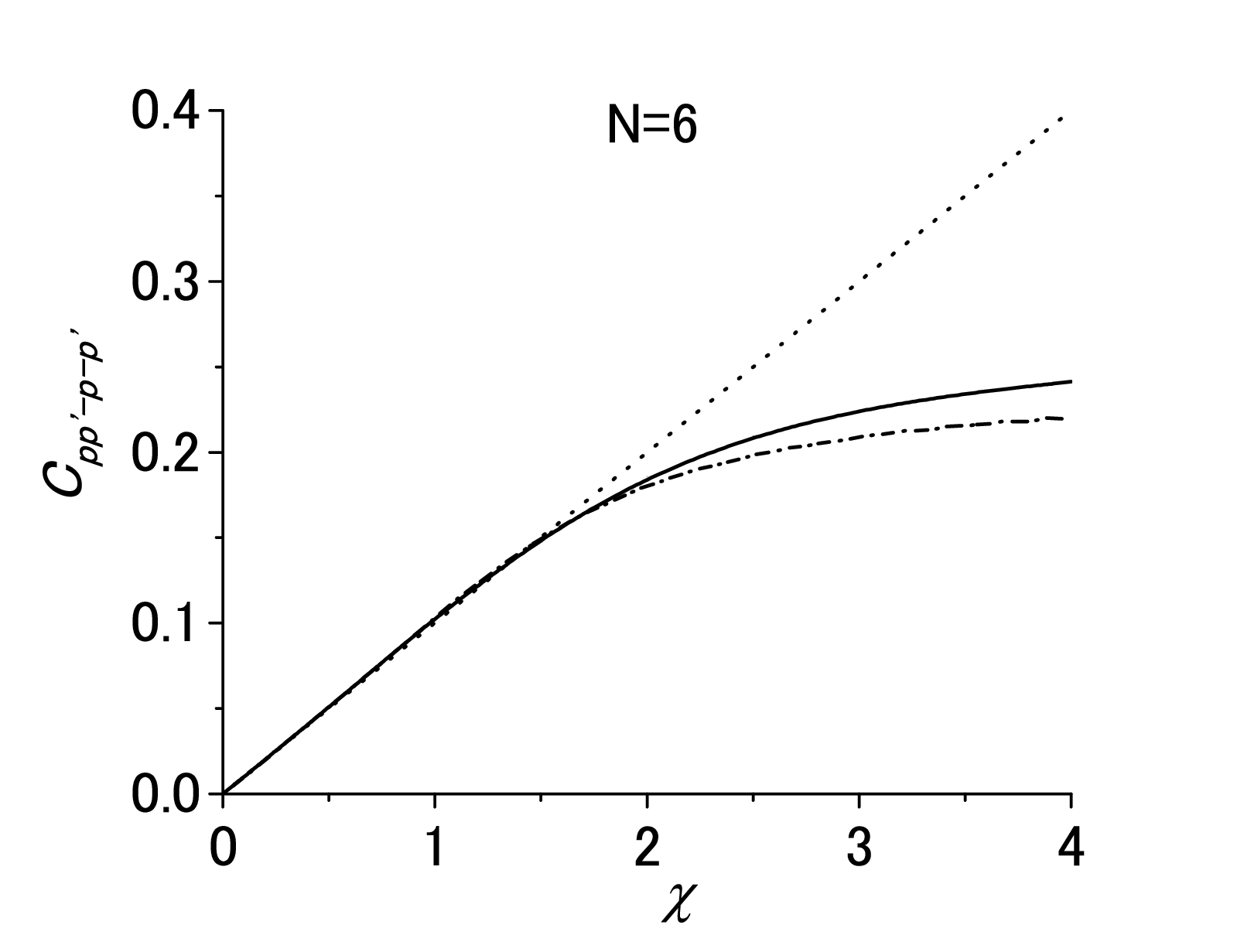}
}
\caption{Same as fig. \ref{pert4C} but for $N=6$.} 
\label{pert6C} 
\end{figure} 
\begin{figure} 
\resizebox{0.5\textwidth}{!}{
\includegraphics{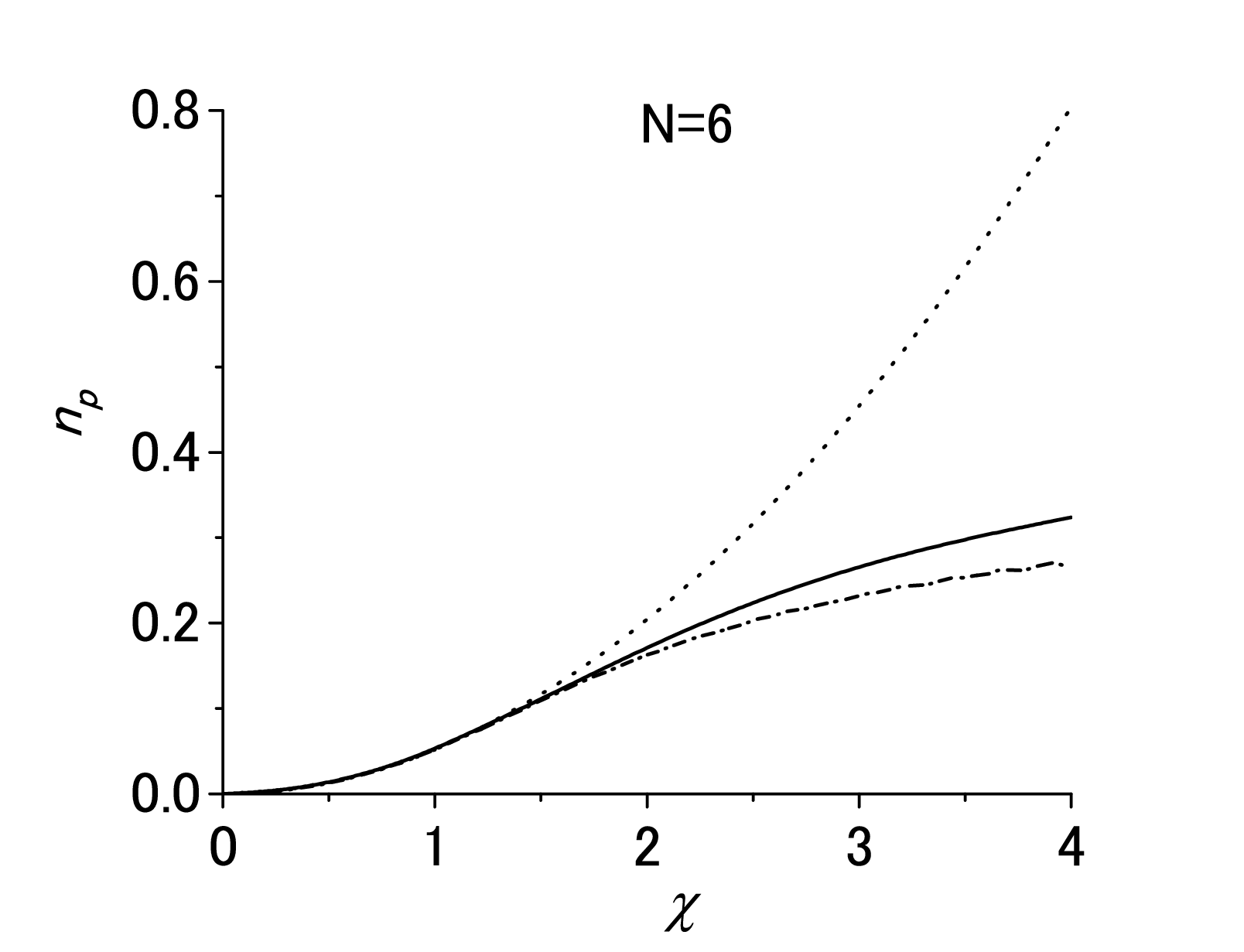}
}
\caption{Same as fig. \ref{pert4n} but for $N=6$.} 
\label{pert6n} 
\end{figure} 
\begin{figure} 
\resizebox{0.5\textwidth}{!}{ 
\includegraphics{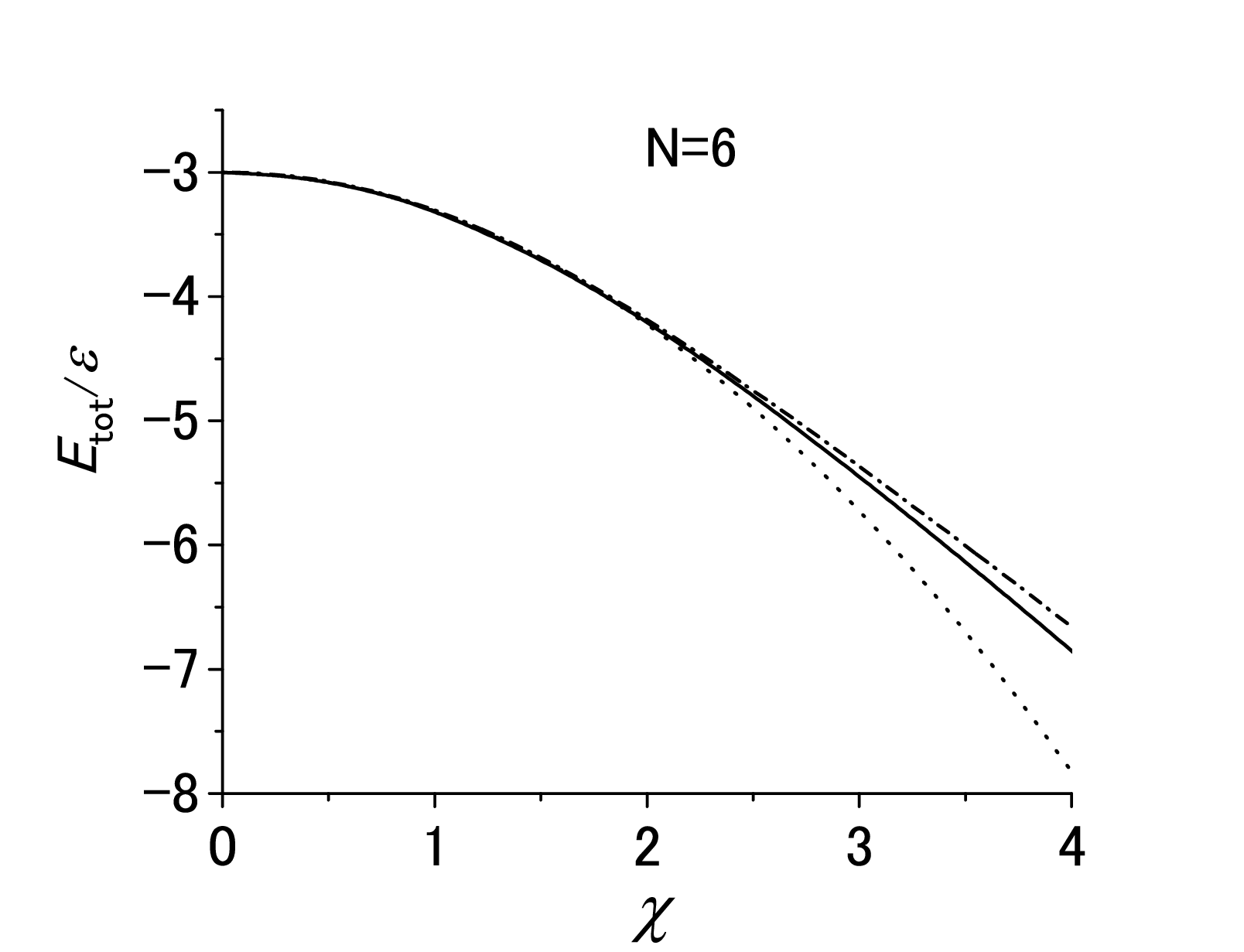}
}
\caption{Same as fig. \ref{pert4E} but for $N=6$.} 
\label{pert6E} 
\end{figure} 
\begin{figure}
\resizebox{0.5\textwidth}{!}{ 
\includegraphics{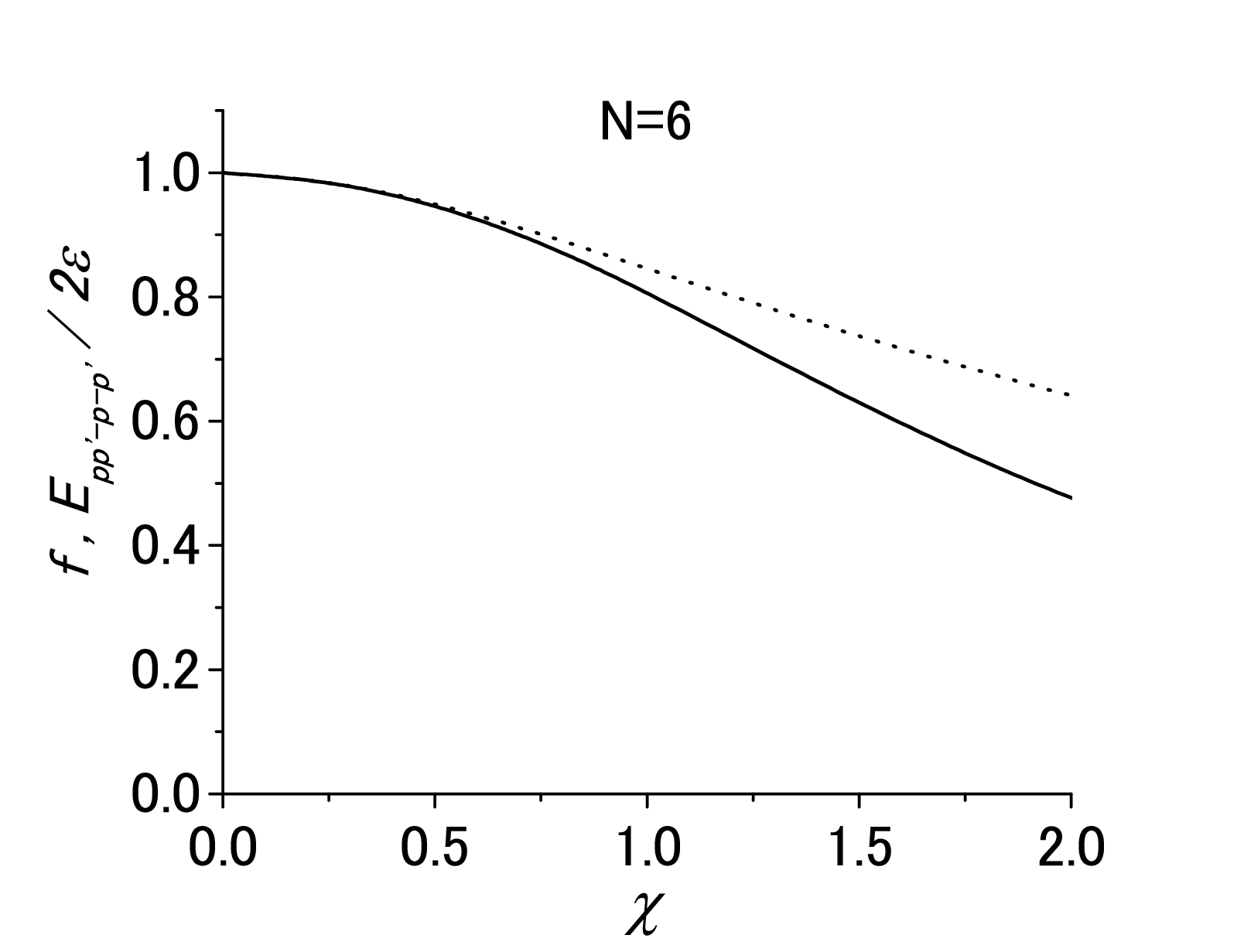}
}
\caption{Occupation factor $f=(1-n_p)(1-n_{p'})n_{-p}n_{-p'}$ (slid line) and the excitation energy $E_{pp'-p-p'}$ of the 2p--2h configuration (dotted line) as functions of $\chi$ for $N=6$.} 
\label{pert6ratio} 
\end{figure} 

The 2p--2h component $C_{pp'-p-p'}$ of $C_2$, the occupation probability $n_p$ of the upper state and the ground-state energy $E_{\rm tot}$ 
calculated in PA (dotted line) and TDDM (dot-dashed line) are shown, respectively, 
in figs. \ref{pert6C}--\ref{pert6E} as functions of $\chi$ for $N=6$. The solid lines depict the exact values. The TDDM results are in good agreement with the exact solutions.
In the case of $N=6$ the agreement of the PA results with the exact solutions extend to $\chi\approx 2$. At $\chi=2$ PA gives $n_p=0.2$ and ${\mathcal N}=2.2$ which mean that it is far beyond the
perturbative region. However, the error of $n_p$ at $\chi=2$ is only 16 \%.
The decrease in the occupation factor is somewhat compensated by the decrease in the energy of the 2p--2h configuration as shown in fig. \ref{pert6ratio} where 
$(1-n_p)(1-n_{p'})n_{-p}n_{-p'}$ and $E_{pp'-p-p'}$ estimated from the exact energy of the second excited state are shown as functions of $\chi$.
The fact that PA gives good results does not mean its validity beyond the perturbative region. It is a result of cancellation between the two competing non-linear effects as shown in fig. \ref{pert6ratio}.

\begin{figure} 
\resizebox{0.5\textwidth}{!}{
\includegraphics{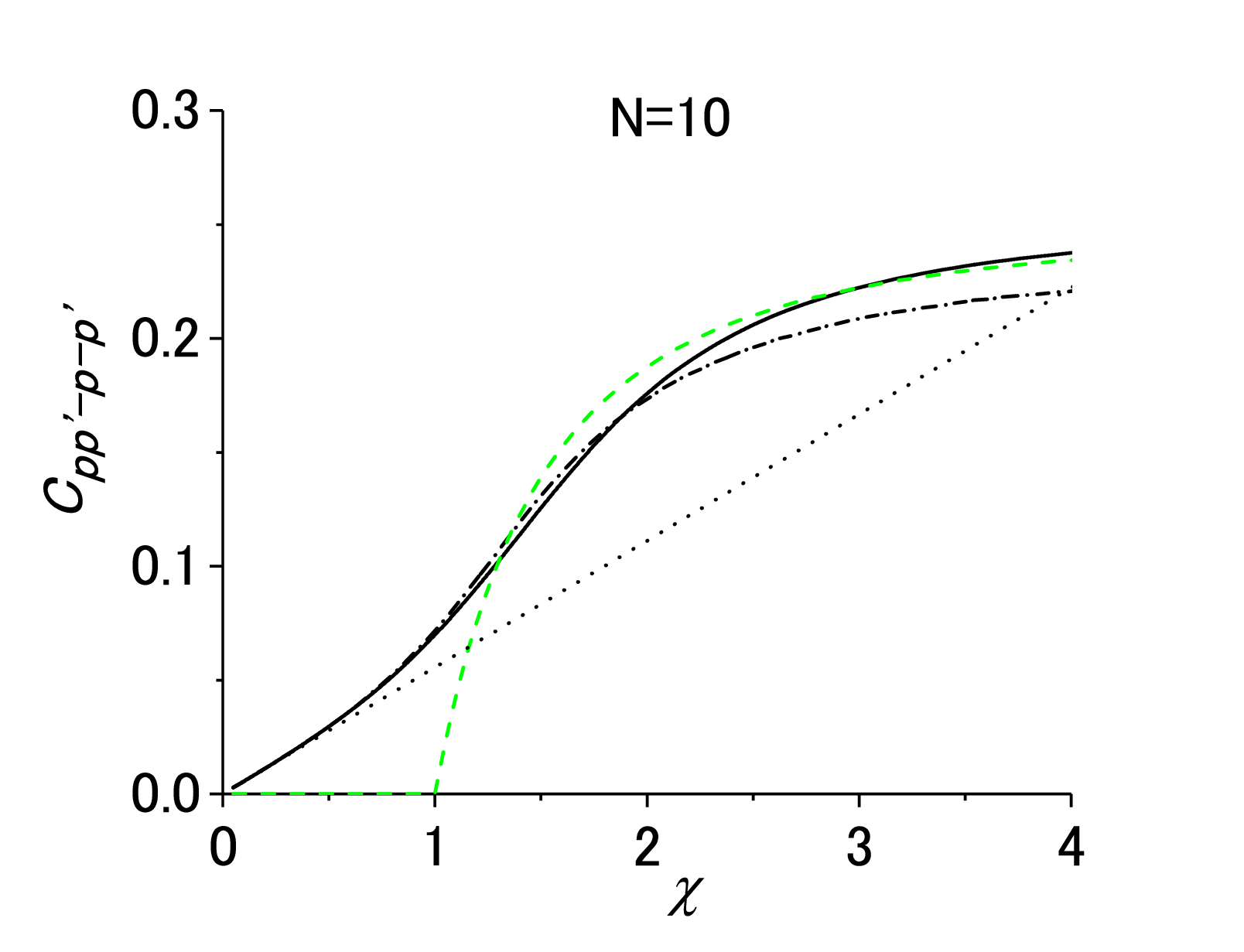}
}
\caption{Same as fig. \ref{pert4C} but for $N=10$. The dashed line depicts the results in DHF.} 
\label{pert10C} 
\end{figure} 
\begin{figure}
\resizebox{0.5\textwidth}{!}{ 
\includegraphics{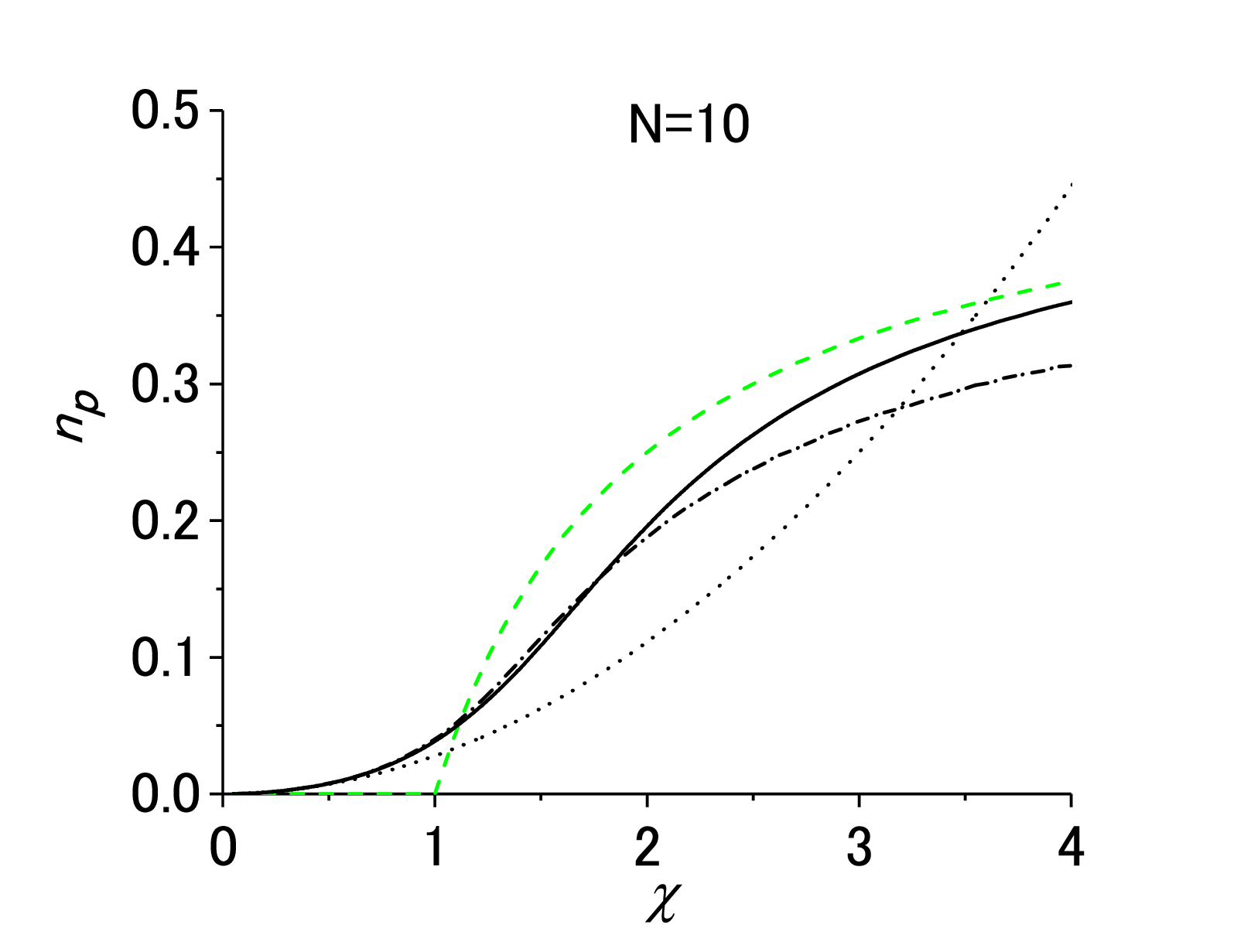}
}
\caption{Same as fig. \ref{pert4n} but for $N=10$. The dashed line depicts the results in DHF.} 
\label{pert10n} 
\end{figure} 

\begin{figure} 
\resizebox{0.5\textwidth}{!}{
\includegraphics{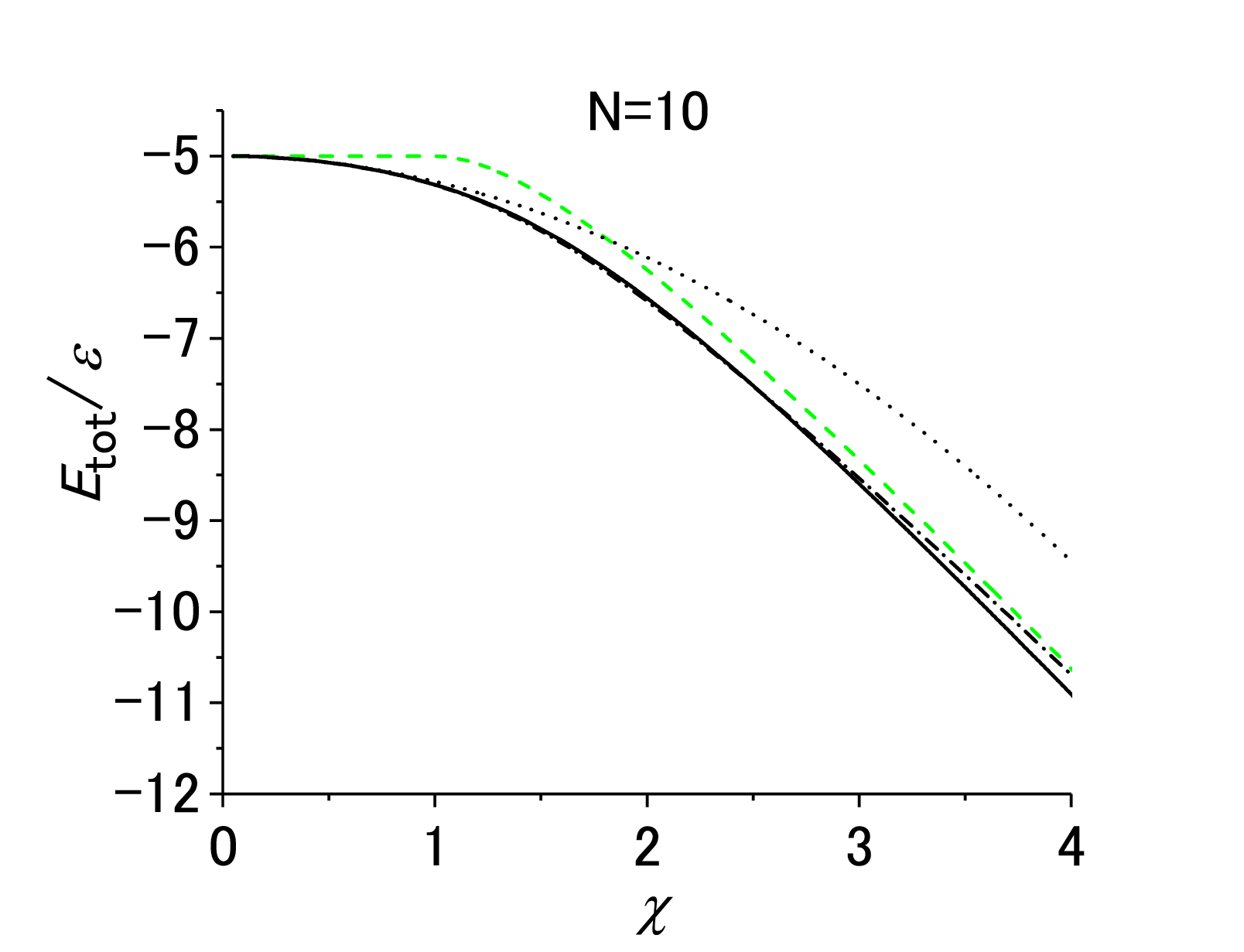}
}
\caption{Same as fig. \ref{pert4E} but for $N=10$. The dashed line depicts the results in DHF.} 
\label{pert10E} 
\end{figure}

\begin{figure}
\resizebox{0.5\textwidth}{!}{ 
\includegraphics{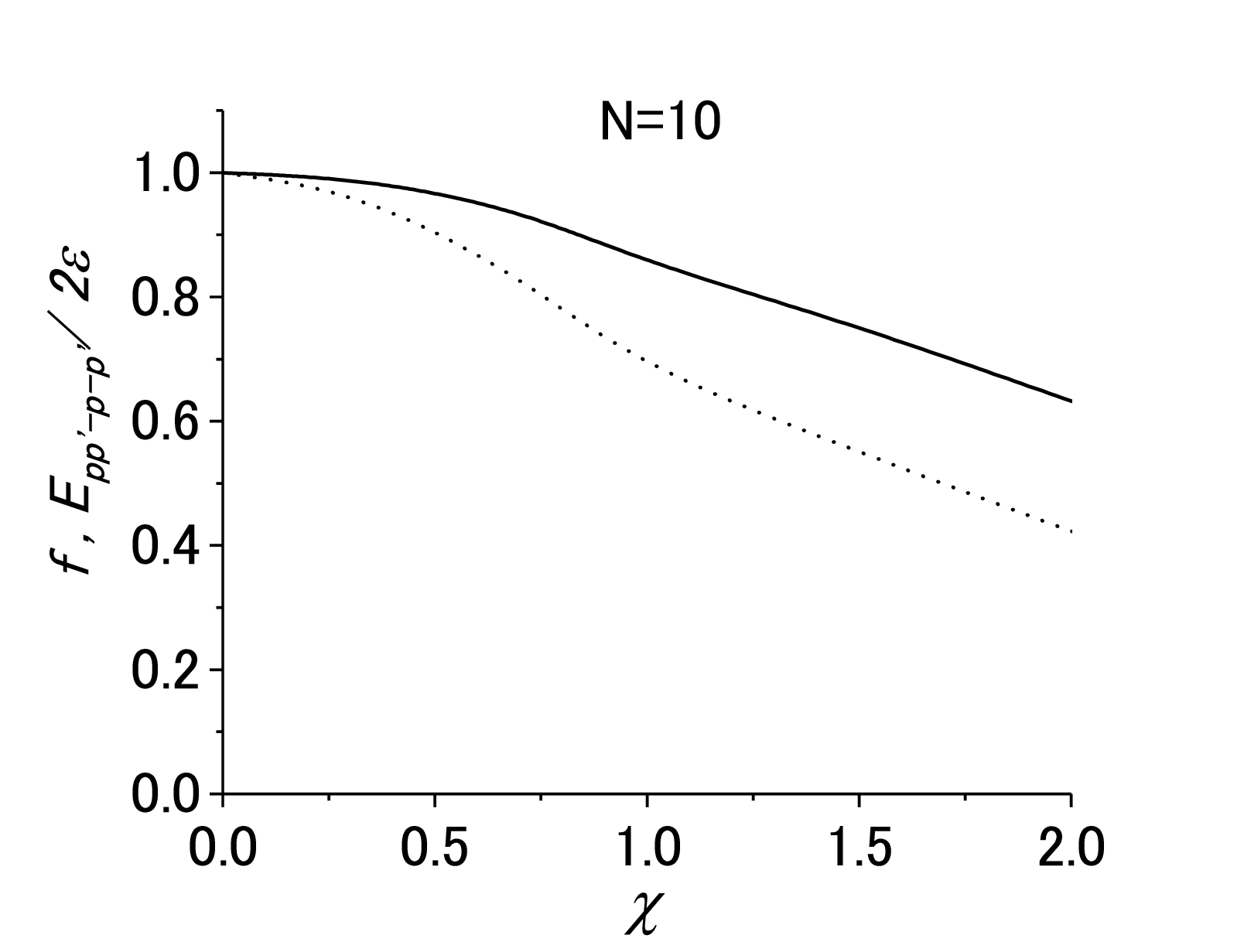}
}
\caption{Same as fig. \ref{pert6ratio} but for $N=10$.} 
\label{pert10ratio} 
\end{figure}

For $N\ge 8$ the validity of PA is again restricted to the perturbative region $\chi <0.5$ as shown in figs. \ref{pert10C}--\ref{pert10E} for $N=10$ where 
the results in DHF are also given with the dashed lines. PA gives ${\mathcal N}<1.07$ for $\chi<0.5$. 
As shown in fig. \ref{pert10ratio}, 
the decrease in the occupation factor $f$ is not compensated by the decrease in the energy of the 2p--2h configuration $E_{pp'-p-p'}$.
The large deviation of the PA results from the exact values beyond $\chi=1$ is related to the fact that $C_{pp'-p-p'}$ does not linearly increase with $\chi$ (see fig. \ref{pert10C})
due to large correlation effects.
It is known that DHF becomes good approximation for large $\chi$ and  $N$ \cite{ts2017},
as shown in figs. \ref{pert10C}--\ref{pert10E}.
At $\chi=1$ the value of $n_p$ is 0.028, the error of $n_p$ is 28 \% and ${\mathcal N}=1.278$.
The errors of PA to this extent might be acceptable.
Thus it is found in the Lipkin model that PA gives good results beyond the perturbative region (${\mathcal N}\approx 1$) in the limiting case of $N=6$
and that in other cases the errors of $n_p$ in PA stay less than 30 \% for $\chi <1$.


\begin{table}
\caption{Proton single-particle energies $\epsilon_\alpha$ and occupation probabilities 
$n_{\alpha}$ calculated in PA, TDDM and EDA for $^{16}$O. }
\begin{center}
\begin{tabular}{c rr rr} \hline
&\multicolumn{1}{c}{}&\multicolumn{2}{c}{$n_{\alpha}$}\\ \hline 
orbit & $\epsilon_\alpha$ [MeV] & PA  & TDDM & EDA  \\ \hline
$1p_{3/2}$ & -17.24 & 0.912 &0.890& 0.888\\
$1p_{1/2}$ & -11.19 & 0.894 &0.865  & 0.856  \\
$1d_{5/2}$ & -3.33 & 0.094 &0.119  & 0.123   \\\hline
\end{tabular}
\label{tab1}
\end{center}
\end{table}

Now let us turn to a realistic case of $^{16}$O.
We performed the PA calculations for $n_\alpha$ using a simple residual interaction \cite{toh07} consisting of only the $t_0$ and $t_3$ terms of the Skyrme III force \cite{skIII}
and the $1p_{3/2}$, $1p_{1/2}$ and $1d_{5/2}$ single-particle states for both protons and neutrons. The results are compared with the results in TDDM and 
in exact diagonalization approach (EDA) obtained
using the same interaction and single-particle states as those used in the PA calculations. The obtained occupation probabilities  
for the proton single-particle states are tabulated in Table \ref{tab1}.
The neutron occupation probabilities only differ from the proton values in the last digit and are not shown.
The TDDM results are in good agreement with the ERA results.
The value of ${\mathcal N}=2.128$ obtained from $n_\alpha$ in PA far exceeds unity, indicating that the interaction strength is far beyond the perturbative region.
However, the errors are not very large:
The errors of $n_\alpha$ for the $1p_{3/2}$, $1p_{3/2}$ and $1d_{5/2}$ states are 21 \%, 26 \% and 24 \%, respectively.
The situation seems similar to the $N=6$ case of the Lipkin model but it is not the case.
In fact the interaction strength is so strong that the EDA energy of the the 1st $0^+$ excited state which mostly consists of 2p--2h configurations 
becomes by $4.42$ MeV lower than the unperturbed HF energy (the ground-state energy in EDA is by 16.27 MeV lower than the energy of the HF configuration).
This makes the PA assumption of the HF ground state and 
the interpretation of the EDA or TDDM results using eq. (\ref{BoE}) meaningless.
For example, $E_{\rm pp'hh'}$ (eq. (\ref{BoE})) for the $1p_{1/2}$ and $1d_{5/2}$ states estimated from the EDA energy of the 1st $0^+$ excited state is negative and its value is given
by $-0.28\times(\epsilon_{\rm p}+\epsilon_{\rm p'}-\epsilon_{\rm h}-\epsilon_{\rm h'})$.
Thus, it is found that in the case of $^{16}$O the perturbative expressions (eqs. (\ref{pertp}) and (\ref{perth})) give 
the occupation probabilities within the accuracy of a few ten percent
in spite of the fact that the perturbation assumptions are badly violated.

 
\section{Summary}
The validity of the perturbative approach (PA) for the occupation probability $n_\alpha$ and the two-body correlation matrix $C_2$ was tested for the Lipkin model by
comparing with the exact solutions. It was found that in the case of $N=6$ PA rather well simulates the exact solutions even in an interaction region where 
perturbative assumptions are not justified. By using the equation for $C_2$ in the time-dependent density-matrix theory (TDDM) it was pointed out that this is a result of cancellation of the decrease in
$(1-n_{\rm p})(1-n_{\rm p'})n_{\rm h}n_{\rm h'}$ with increasing (decreasing) $n_{\rm p}$ ($n_{\rm h}$) and that in the energy of the 2p--2h configurations with increasing interaction strength.
It was also found that for $N=4$ and $N\ge 8$ the validity of PA is limited to a genuine perturbative region ($\chi < 0.5$). 
PA was also applied to $^{16}$O and it was found that the occupation probabilities in PA are not very different from the results in exact diagonalization approach though 
perturbation assumptions are not fulfilled.

\end{document}